\documentclass[usenatbib]{mn2e}

\usepackage{graphicx}
\usepackage{amssymb}

\newcommand{\apj}{ApJ}
\newcommand{\apjl}{ApJ}
\newcommand{\apjs}{ApJS}
\newcommand{\aj}{AJ}
\newcommand{\aap}{A\&A}
\newcommand{\mnras}{MNRAS}

\newcommand{\apss}{Ap\&SS}

\newcommand{\pasp}{PASP}

\newcommand{\arXiv}{arXiv}
\newcommand{\araa}{ARA\&A}

\newcommand{\integral}{INTEGRAL}
\newcommand{\swift}{\emph{Swift}}
\newcommand{\chandra}{\emph{Chandra}}
\newcommand{\spitzer}{\emph{Spitzer}}
\newcommand{\planck}{\emph{Planck}}
\newcommand{\glimpse}{GLIMPSE}
\newcommand{\mipsgal}{MIPSGAL}

\newcommand{\igr}{IGR\,J17448-3232}
\newcommand{\cxopt}{CXOU\,J174437.3-323222}
\newcommand{\cxoex}{CXOU\,J174453.4-323254}

\newcommand{\swft}{Swift\,J174437.5-323220}
\newcommand{\pbc}{2PBC\,J1744.8-3231}
\newcommand{\glimp}{G356.8134-01.6971}

\newcommand{\ergscm}{erg\,cm$^{-2}$\,s$^{-1}$}

\newcommand{\arcdeg}{$^{\circ}$}

\title[Discovery of the IR counterpart of \cxopt]
{Discovery of the infrared counterpart of \cxopt\ in the field of \igr:\\ a blazar candidate viewed through the Galactic centre?}

\author[P.A.~Curran et al.]
{P.A.~Curran$^1$\thanks{e-mail: peter.curran@cea.fr},
S.~Chaty$^{1}$,
J.A.~Zurita Heras$^{2}$,
J.A.~Tomsick$^{3}$,
T.J.~Maccarone$^{4}$
\\
$^1$Laboratoire AIM, CEA/IRFU-Universit\'e Paris Diderot-CNRS/INSU, CEA DSM/IRFU/SAp, Centre de Saclay, F-91191  Gif-sur-\\ Yvette, France \\
$^2$Fran\c{c}ois Arago Centre, APC, Universit\'e Paris Diderot, CNRS/IN2P3, CEA/DSM, Observatoire de Paris, 13 rue Watt, 75205\\ Paris Cedex 13, France \\
$^3$Space Sciences Laboratory, 7 Gauss Way, University of California, Berkeley, CA 94720-7450, USA\\
$^4$School of Physics and Astronomy, University of Southampton, Southampton, Hampshire, SO17\,1BJ, UK \\
}

\begin{document}

\date{Accepted/ Received;}

\pagerange{\pageref{firstpage}--\pageref{lastpage}} \pubyear{}

\maketitle

\label{firstpage}


\begin{abstract}
We present our near infrared ESO-NTT $K_S$ band observations$^1$ of the field of \igr\ which show no extended emission consistent with the SNR but in which we identify a new counterpart,  also visible in \spitzer\ images up to 24\,$\mu$m,  at the position of the X-ray point source, \cxopt.  
Multi-wavelength spectral modelling shows that the data are consistent with a reddened and absorbed single power law over five orders of magnitude in frequency. This implies non-thermal, possibly synchrotron emission that renders the previous identification of this source as a possible pulsar, and its association to the SNR, unlikely;  we instead propose that the emission may be due to a blazar viewed through the plane of the Galaxy. 
\end{abstract}



\section{Introduction}\label{section:introduction}

\let\thefootnote\relax\footnotetext{\noindent$^1$ Based on observations collected at the European Organisation for Astronomical Research in the Southern Hemisphere, Chile under ESO program 084.D-0535 (P.I. Chaty)}

High energy (X-ray, $\gamma$-ray) emission is observed from numerous astronomical sources and is produced by various processes. Common sources of high energy emission are Galactic sources such as black hole or neutron star systems (e.g., X-ray binaries, pulsars) and supernovae or supernova remnants, and extragalactic sources such as active galactic nuclei (AGN) and gamma-ray bursts (GRBs). At lower energies ($\lesssim 20$\,keV), thermal emission may make a significant contribution to the observed flux but at greater energies, such as those observed by the \emph{International Gamma-Ray Astrophysics Satellite} (\integral;  15\,keV - 10\,MeV; \citealt{Winkler2003:A&A411L}), other emission mechanisms are required. Prominent, though certainly not the only emission mechanisms at these energy ranges are synchrotron and inverse Compton radiation which both produce broad-band spectra visible over many orders of magnitude in frequency. 

In this Letter we collate data of one such \integral\ detected, high energy object, \igr,  from various sources including published catalogs, archived images and our own ESO-NTT observations (section \ref{section:observations}). We identify the multi-wavelength counterparts of the point source and construct a spectral energy distribution (SED; section \ref{section:spectra}) in an attempt to understand its nature.

\noindent {\bf \igr}\label{section:introduction-igr}\\
\noindent 
The X-ray source \igr\ was initially discovered by INTEGRAL
and published in the Third (III) and subsequently Fourth (IV) IBIS/ISGRI Soft Gamma-Ray Survey Catalog \citep{Bird2007:ApJS170,Bird2010:ApJS186}, though at slightly different positions.
In an attempt to refine the position (see Table\,\ref{table:xrays} for this and subsequent X-ray positions), the \emph{Swift} X-ray telescope (XRT; \citealt{burrows2005:SSRv120}) observed the field \citep{Landi2007:ATel.1323}. These authors identify a point source (henceforth \swft) at the edge of the original \integral\ III error circle as well as possible diffuse emission (which, after examination of the XRT image, we note is within that error circle). 
The position of the point source was further refined by \cite{Tomsick2009:ApJ701} through  a 4.7\,ks  \chandra\ observation of the field. In addition to detecting the point source, \cxopt, 
they confirmed an extended source, $\sim7$ times brighter, at the \integral\ III position (\cxoex, see figure\,\ref{fig:24um}). 
At the XRT point source position, \cite{Landi2007:ATel.1323} had also noted a USNO-B1.0 (0574-0773466, $R \sim 15.5$; \citealt{Monet2003:AJ125}), 2MASS (J17443749-3232197, $K=9.100 \pm 0.026$; \citealt{Skrutskie2006:AJ.131}) object but the sub-arcsecond accuracy of the \chandra\ position eliminates it as a possible counterpart; though the X-ray point source is at the edge of the 2MASS point spread function (PSF).

Due to the fact that the more recent \integral\ IV position includes both the point source and extended emission within its error circle we shall henceforth only use \igr\ to refer to the field and, for clarity, we shall refer to the extended emission as \cxoex. Given the positional coincidence and point source nature of \swft\ 
 and \cxopt\ we shall henceforth assume they are one source which we shall refer to by the \chandra\ moniker. 
Based on analysis of the \chandra\ spectrum, \cite{Tomsick2009:ApJ701} proposed the extended emission as a supernova remnant (SNR), though they did not detect evidence of a pulsar wind nebula (PWN). They also suggested a tentative association of the point source with the SNR, hypothesizing that it may be an isolated neutron star that received a kick when the supernova occurred.

\begin{table}	
  \centering	
  \caption{X-ray positions and 90\% uncertainties for the different X-ray sources in the field.} 	
  \label{table:xrays}
  \begin{tabular}{l l l l}
    \hline\hline
    Source & RA & Declination & Error \\
    \hline 
    \igr$^{\rm{III}}$   & 17:44:54.96 & -32:33:00 & 2.2\arcmin \\
    \igr$^{\rm{IV}}$   & 17:44:47.76 & -32:32:16 & 3.2\arcmin \\
    \swft$^{*}$  & 17:44:37.23 & -32:32:24.3 &  2.4\arcsec \\

    \cxoex$^e$ & 17:44:53 & -32:32:54 &   \\
    \cxopt & 17:44:37.34 & -32:32:23.0 &  0.64\arcsec \\
    \pbc   & 17:44:52.6  & -32:31:01   &   4.56\arcmin \\

    \hline 
  \end{tabular}
  \begin{list}{}{}
  \item[] 
    $^{\rm{III/IV}}$ Position from the Third/Fourth IBIS/ISGRI Soft Gamma-Ray Survey Catalog.
    $^*$ Refined in this paper from  original position of 17:44:37.46 -32:32:20.2 (4\arcsec\ error; \citealt{Landi2007:ATel.1323}).
    $^e$ Extended source of radius $\sim$ 3\arcmin.
  \end{list}
\end{table}


\section{Observations}\label{section:observations}

\subsection{Catalog sources}\label{catalogs}

In the high energy range ($\gtrsim 10$\,keV) a source, \pbc, is documented at  $1.2\times 10^{-11}$ \ergscm\
in the Palermo Swift-BAT hard X-ray catalogue (15--150\,keV; \citealt{Cusumano2010:A&A524}). The source is therein associated with \igr, though its position is consistent with both the extended and point sources in the field; given the resolution of the BAT instrument this is an unresolved measurement of the flux at that position. 
No source, consistent with either position, is found in the \emph{Fermi} LAT First Source Catalog (100\,MeV -- 100\,GeV; \citealt{Abdo2010:ApJ188}) and no such source has been made public by \emph{Fermi} GBM (10\,keV -- 30\,MeV) or HESS (100\,GeV -- 100\,TeV; \citealt{Chaves2009:arXiv0907}).

The second epoch Molonglo Galactic Plane Survey (MGPS-2) compact source catalogue \citep{Murphy2007:MNRAS382}, which details observations at 843\,MHz, documents two nearby sources, MGPS\,J174442-323433 and MGPS\,J174455-323357, which are also included in the NRAO VLA Sky Survey (1.4\,GHz; NVSS; \citealt{Condon1998:AJ115}) as NVSS\,J174442-323436 and NVSS\,J174455-323359. 
The extension of these sources overlap with the extended X-ray emission of the SNR but neither are consistent with the point-like X-ray source. From a visual inspection of the NVSS and  MGPS-2 images available of the field, there is no obvious sign of any excess emission above background levels at the point source position; we use the flux density of the nearby dim, NVSS\,174423-323502, as a measure of the 1.4\,GHz upper limit in the field ($2.8 \times 10^{-3}$\,Jy).
No nearby sources are found in any of the 9 bands (30\,GHz -- 856\,GHz) of the all-sky, \planck\ Early Release Compact Source Catalog \citep{Planck2011:arXiv1101}; flux density limits for the region are not well quantified but are on the order of 1\,Jy for latitudes up to $\pm10$\arcdeg\ (\citealt{Planck2011:arXiv1101} figure\,5).

\begin{figure}
  \centering 
  \resizebox{\hsize}{!}{\includegraphics[angle=-0]{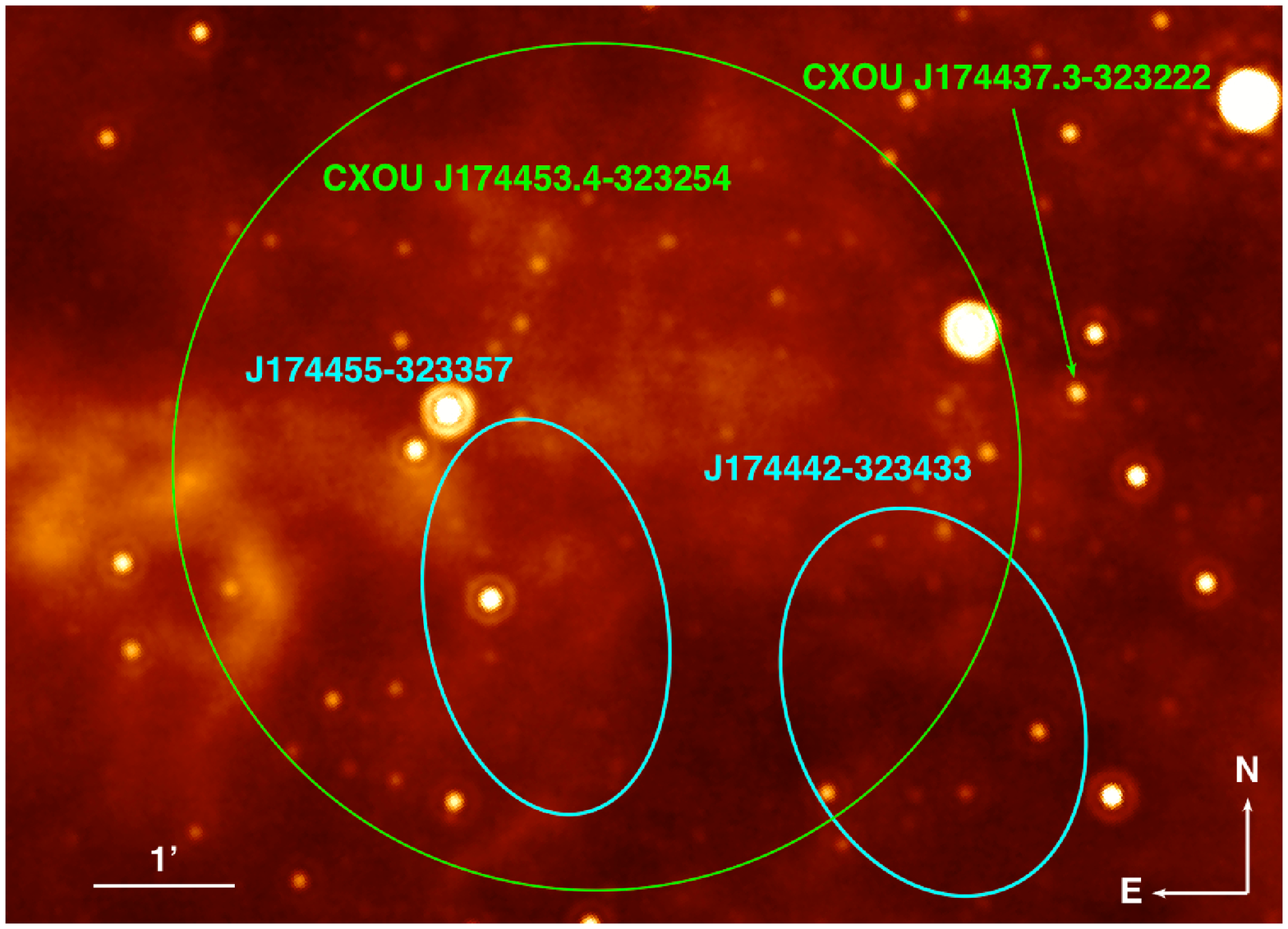} }
  \caption{24$\mu$m MIPSGAL image of the field of \igr\ showing the blazar candidate, \cxopt, as well as the extent of the $\sim 3\arcmin$ extended X-ray emission from the SNR and the extended MGPS-2 radio sources. }
  \label{fig:24um} 
\end{figure}

\subsection{Near IR observations}\label{sofi}

\begin{figure}
  \centering 
  \resizebox{\hsize}{!}{\includegraphics[angle=-0]{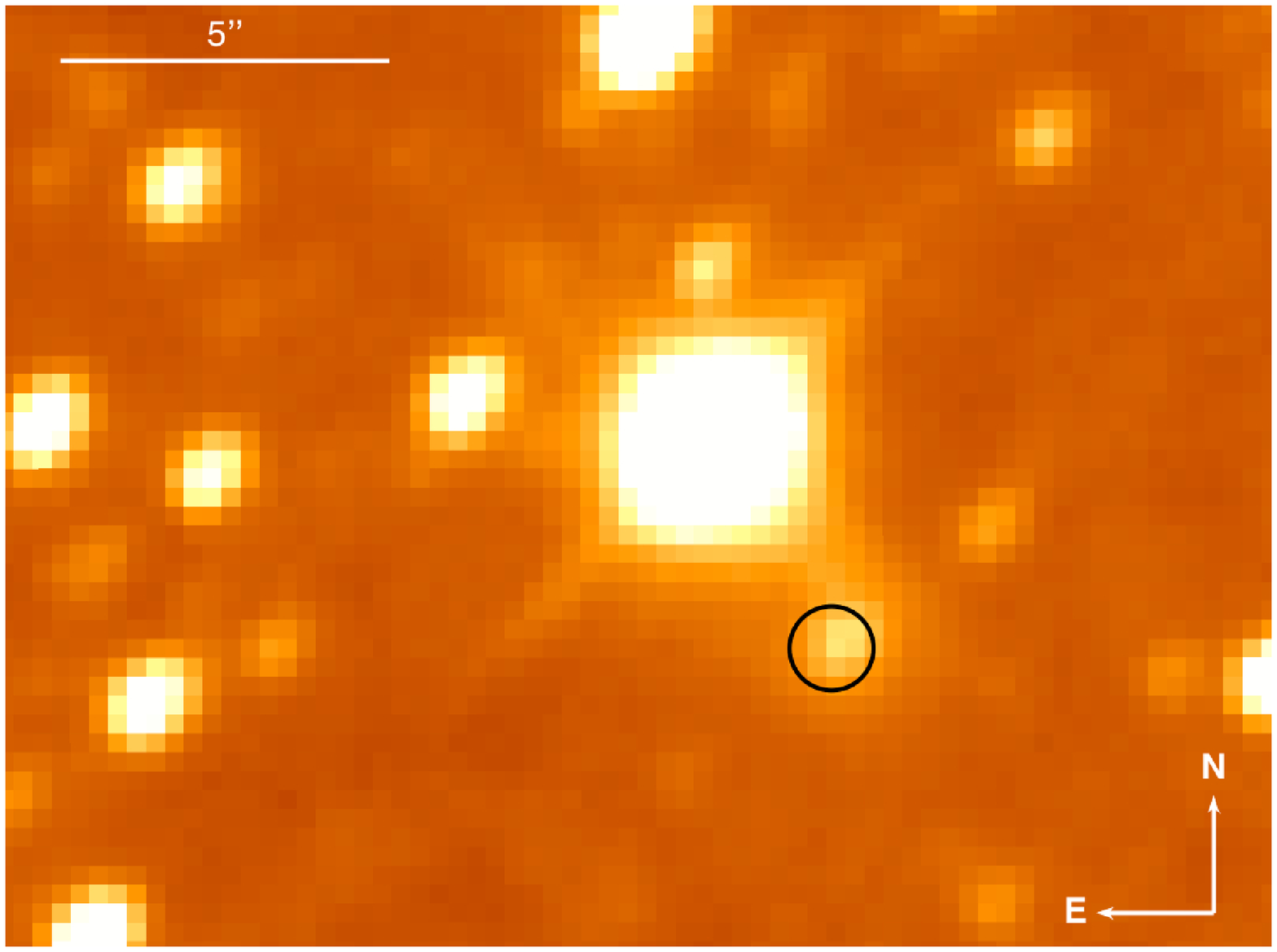} }
  \caption{SofI $K_{S}$ band 20\arcsec $\times$ 15\arcsec\ image with \chandra\ 0.64\arcsec\ 90\% error circle marked.}
  \label{fig:Ks} 
\end{figure}

On March 28, 2010 a total of 9 10-second $K_{S}$ filter exposures  were obtained with the Son of ISAAC (SofI) infrared spectrograph and imaging camera on the 3.58m ESO-New Technology Telescope (NTT). The NTT-SofI data were reduced using the {\small IRAF} package wherein crosstalk correction, flatfielding, sky subtraction and frame addition were applied. 
The image was astrometrically  calibrated against 2MASS \citep{Skrutskie2006:AJ.131} within the GAIA package. 
No extended emission consistent with the diffuse X-ray emission is found in the image, but a new source is detected at the position of the \chandra\ point source (Figure\,\ref{fig:Ks}). 
The $K_{S}$ counterpart is at the edge of, and has an overlapping PSF with, the aforementioned 2MASS source and hence PSF photometry is required to accurately determine the magnitude.

PSF photometry was carried out on the final image using the {\small DAOPHOT} package \citep{stetson:1987PASP} within {\small IRAF}. The magnitude of the counterpart was then calculated relative to a number of comparison stars in the field, using the scatter as a measure of the error. The comparison stars were calibrated against a Persson photometric standard star \citep{Persson1998:AJ116} observed on the night. The resultant magnitude and $1 \sigma$ error of the nIR source is $K_S = 14.03 \pm 0.20$, corresponding to a flux of (1.63 $\pm$0.33)$\times 10^{-3}$\,Jy (see also Table\,\ref{obs}). The source appears point like, with no evidence of extended emission. 
The approximate probability of a chance superposition down to the observed magnitude of the source is 3\%, 
or down to the limiting magnitude of the field ($K_{S} =19.0$) is 11\%; these are the number densities of observed sources in the field at those magnitudes, times the area of the X-ray positional error.

\subsection{\swift\  observations}

We are further able to refine the \swift\ XRT position of \cite{Landi2007:ATel.1323}, using the XRT online tool
\citep{evans2009:MNRAS397,Goad2007:A&A476}. This position (Table\,\ref{table:xrays}), based on a 3.6\,ks exposure,   
 is consistent with the previous XRT and \chandra\ positions as well as our newly proposed counterpart, but excludes the 2MASS source. 
While \swift\ was observing the field of \igr, it was also observing with the Ultraviolet/Optical Telescope (UVOT; \citealt{roming2005:SSRv120}). From the \swift\ archive we find that 3592\,s of $uw2$ image data were obtained. Using {\tt FTOOLS} we summed the image data  and, since no source was found at the position, derived a limiting magnitude of 21.82 (UVOT photometric system; \citealt{poole2008:MNRAS383}), equivalent to  $1.38 \times 10^{-6}$ Jy at $1.477 \times 10^{15}$ Hz.

\subsection{\spitzer\  observations}\label{spitzer}

We utilise data from the \emph{Spitzer Space Telescope}'s  \citep{Werner2004:ApJS154} \glimpse\ and \mipsgal\ surveys. The Galactic Legacy Infrared Mid-Plane Survey Extraordinaire (\glimpse; \citealt{Benjamin2003:PASP115}) was carried out by the IRAC instrument \citep{Fazio2004:ApJS154} aboard \spitzer. \glimpse\ spans 65\arcdeg\ either side of the Galactic center up to $\pm$2-4\arcdeg\ in latitude and includes mosaic images and catalog entries at 3.6, 4.5, 5.8 and 8.0 $\mu$m. At the X-ray point source position there is a \glimpse\ cataloged source, \glimp, as well as one coincident with the nearby 2MASS source. Two consistent magnitudes are given at each wavelength so we calculate a weighted averaged and error for each (Table\,\ref{obs}). 
\mipsgal\ \citep{Carey2009:PASP121} is a survey of the inner Galactic plane using the MIPS instrument \citep{Rieke2004:ApJS154} aboard \spitzer. It covers approximately the same longitudes as \glimpse, up to latitudes of $\pm1$\arcdeg, at 24 and 70  $\mu$m, though at the moment only the 24\,$\mu$m mosaic product (Figure\,\ref{fig:24um}) has been released and no catalog is available.  A source coinciding with the X-ray point source is visible in this image though the  2MASS source, visible at shorter wavelengths, is not. We have derived the flux and magnitude of this source by aperture photometry in {\small IRAF} and the photometric information in the image header (Table\,\ref{obs}).  
There is no sign of any diffuse emission corresponding to the extended X-ray emission in any of the \glimpse\ or \mipsgal\ mosaic images available.

\begin{table}	
  \centering	
 \caption{Observed magnitudes and flux densities, $F_{\nu}$, for IR counterparts of \cxopt\ ($1\sigma$ errors).}
  \label{obs}
  \begin{tabular}{l l l} 
    \hline\hline
    Band (Instrument) & Magnitude & Flux Density (Jy) \\ 
    \hline  
    UV (\swift)	        &  $>$ 21.82      &  $< 1.38 \times 10^{-6}$  \\ 
    $K_S$ (SofI)	&  14.03 $\pm$ 0.20      &  (1.63 $\pm$ 0.16)$\times 10^{-3}$  \\ 
    3.6$\mu$m (IRAC)	&  12.66 $\pm$ 0.11      &  (2.42 $\pm$ 0.27)$\times 10^{-3}$ \\
    4.5$\mu$m (IRAC)	&  12.00 $\pm$ 0.08      &  (2.85 $\pm$ 0.29)$\times 10^{-3}$ \\
    5.8$\mu$m (IRAC)	&  11.18 $\pm$ 0.06      &  (3.88 $\pm$ 0.23)$\times 10^{-3}$ \\
    8.0$\mu$m (IRAC)	&  10.12 $\pm$ 0.04      &  (5.74 $\pm$ 0.23)$\times 10^{-3}$ \\
    24$\mu$m (MIPS)	&   6.7  $\pm$ 0.3       &  (1.5  $\pm$ 0.5)$\times 10^{-2}$
 \\
    \hline 
  \end{tabular}
\end{table}


\section{Multi-wavelength SED}\label{section:spectra}

\begin{figure}
  \centering 
  \resizebox{\hsize}{!}{\includegraphics[angle=-90]{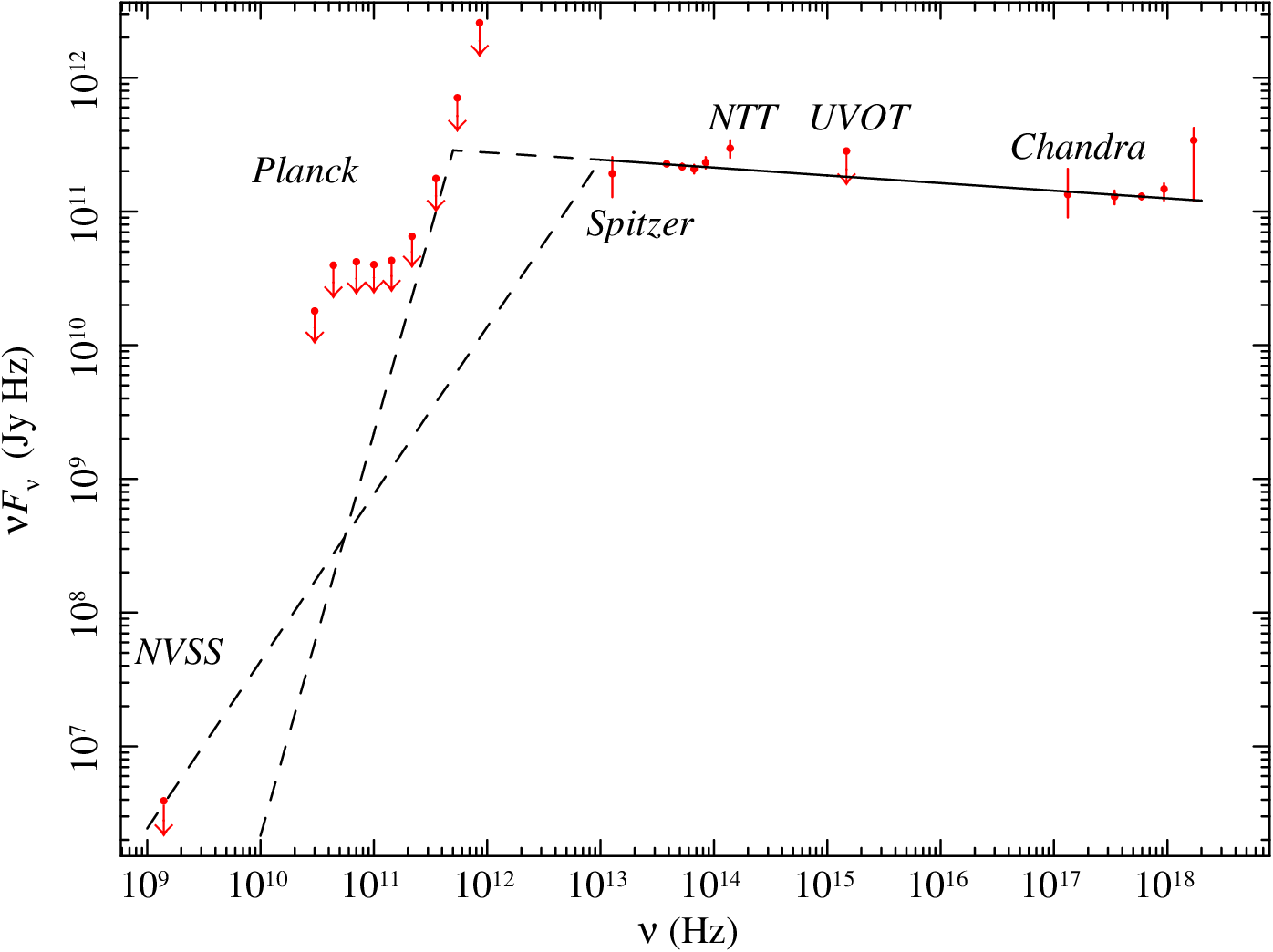}}
  \caption{Unabsorbed/dereddened \citep{cardelli1989:ApJ345} SED for the point source, showing power law fit ($\alpha = -1.057$) and two possible low frequency spectral slopes ($\alpha = 0.25$, $\alpha = 2.0$).} 
  \label{fig:SED} 
\end{figure}

The infrared flux densities in Jansky (Jy), $F_{\nu}$, at frequency $\nu$ (Table\,\ref{obs}) were first converted to 
flux per filter, $F_{filter}$ in units of photons\,cm$^{-2}$\,s$^{-1}$. 
This is done via  $F_{filter} =  1509.18896  F_{\nu}$ $( \Delta\lambda/\lambda ) $ 
where $\lambda$ and $\Delta\lambda$ are the effective wavelength and width of the filter in question. 
{\tt XSPEC} compatible files are then produced using the {\tt FTOOL}, {\tt flx2xsp}. 
Within {\tt XSPEC}, the infrared, \swift\ UVOT and \chandra\ X-ray \citep{Tomsick2009:ApJ701} data were initially fit by a single power law where interstellar extinction and absorption were modeled by {\tt redden} and {\tt phabs} respectively. 
However, this model gave an unsatisfactory fit ($\chi^{2}_{\nu} = 1.82$ for 16 degrees of freedom), namely due to a clear excess of emission at energies $>5$\,keV that could not be accounted for by any possible pile-up. Including an additional, purely phenomenological, power law component produces an improvement ($\chi^{2}_{\nu} = 1.00$) of the fit with power law and extinction/absorption parameters similar to those of the initial fit. 
The best fit (Figure\,\ref{fig:SED}) gives a broadband spectral index, $\alpha$ ($F_{\nu} \propto \nu^{\alpha}$) from infrared to X-ray of $\alpha = -1.057 \pm 0.015$ ($1\sigma$ confidence), as well as a secondary spectral index (not plotted) to account for the excess emission $>5$\,keV which can only be constrained to be within the range $1.4 < \alpha^\prime < 2.3$. 
Unabsorbed X-ray fluxes and flux densities  were calculated in five energy bins: 0.3--1, 1--2, 2--3, 3--5 and 5--10 keV.

The optical extinction is fit as  $E_{(B-V)} = 0.84^{+0.3}_{-0.15}$  and the equivalent hydrogen column density as $N_{\mathrm{H}} = 2.69 \pm 0.25 \times 10^{22}$\,cm$^{-2}$, greater than the Galactic value of  $N_{{\rm H\,Galactic}}  = 0.67 \times 10^{22}$\,cm$^{-2}$ \citep{kalberla2005:A&A440}.  These should be treated with caution due to the small range over which they are calculated and, in the case of extinction, the lack of sensitivity at those wavelengths. 
It is also worth noting that the best fit optical extinction is well below that of the Galactic value, $E_{(B-V)\,\mathrm{Galactic}} = 3.761$  \citep{schlegel1998:ApJ500}, though this should be treated with caution as estimates of the extinction so close to the Galactic plane ($<5$\arcdeg) are unreliable.
The measured column density to the source implies an optical extinction \citep{Guver2009:MNRAS.400} of $E_{(B-V)\,\mathrm{Galactic}} = 3.9 \pm 0.4$, well in excess of the observed value, which suggests an intrinsic excess of $N_{\mathrm{H}}$ at the source. 

The \planck\ limits imply that the spectrum breaks at a frequency, $\nu_{\rm{break}}$: $5 \times 10^{11} \lesssim \nu_{\rm{break}} \lesssim 1 \times 10^{13}$\,Hz, i.e., between the \planck\ and \spitzer\ frequencies, though the exact value will depend on the sharpness of the break. The flattest possible spectral index, $\alpha$ at  low frequencies is $\approx 0.25$. The upper limits are also consistent with a steepest possible (physical) spectral slope of $\alpha=2$.

\section{Discussion}\label{section:discussion}

We detect an IR point source, with no evidence for extended emission, at the X-ray position of \cxopt, the X-ray point source in the field of \igr. The probability of chance superposition, even in this crowded field, is relatively low at 11\% in the $K_S$ band, or only 3\% for a source of this magnitude or brighter, and lower (1-4\%) in the more sparsely populated \glimpse\ fields.
As X-ray flux measurements from \swift-XRT and \chandra\ are consistent, we can assume that the source is relatively persistent, 
though we cannot rule out variability. The IR to X-ray SED displays a single power law, with only a suggestion of excess emission above $5$\,keV (this single power law SED derived from non-contemporaneous archive/catalog values is another argument against a significantly variable source). There is no evidence in the SED for any thermal emission at any frequency in the observed bands. 
The spectral index of the power law, $\alpha \sim -1.0$, could correspond to the expected spectral slope of synchrotron emission from accelerated electrons; such electrons, accelerated to a power-law energy distribution, $\mathrm{d}N \propto E^{-p} \mathrm{d}E$ with a cut-off at low energies, are expected, under standard assumptions, to emit high energy photons with a spectral index of $\alpha = -(p-1)/2$ up to the cooling break frequency, which is due to the finite synchrotron-emitting lifetimes of the electrons, and $\alpha = -p/2$ thereafter (e.g., \citealt{Longair1994:hea.book}).
The value of the electron energy distribution index, $p$, is thought to have a value of $\sim 2.0-2.5$ (e.g., \citealt{kirk2000:ApJ542,achterberg2001:MNRAS328,spitkovsky2008:ApJ682,Curran2010:ApJ716L}), so the observed value of $\alpha \sim -1.0$ could correspond to a value of $p=2.0$ for observations above the cooling frequency, though this is only one possible interpretation of the data.

This persistent, single power law  emission might be expected from a number of astronomical sources such as AGN, persistent X-ray binaries or magnetars but in most of these cases there would be a measurable thermal component, of which there is no evidence of here. Power law emission would also be expected from a SNR or PWN but the extended nature of these sources should be observable in either the X-ray or IR/optical, which is not the case (e.g., \citealt{Gaensler2006:ARA&A.44}).
As previously mentioned above, \cite{Tomsick2009:ApJ701} suggested a tentative association of the point source with the nearby SNR, hypothesizing that it may be an isolated neutron star that received a kick when the supernova occurred. However, a pulsar would not be expected to be so bright in the nIR relative to the radio regime, where a detection would be expected but is not the case here. 
Noting that the source is so close to the Galactic center; if we assume a distance of 8\,kpc, then the angular separation of 3\arcmin\ (corresponding to $\sim8$\,pc) would require $\sim$16,000 years since the super nova (SN), if a pulsar was given a kick velocity of 500\,km\,s$^{-1}$ (at the high end of the distributions of \citealt{Brisken2003:AJ.126,Hobbs2005:MNRAS.360}). At that age we would expect that the SN would have faded significantly which does not seem to be the case, though of course this time can be reduced significantly by reducing the distance to the source. While we cannot rule out an association, we find it unlikely, and to confirm the point source as a pulsar, associated or not to the SNR, X-ray timing analysis is required.


One solution that can explain the emission is that the source is a blazar (e.g., \citealt{Urry1995:PASP.107,Padovani2007:Ap&SS.309,Ghisellini2011:arXiv1104}), an AGN with its jet pointing directly at us. Blazars are persistent sources, though it should  be noted that they have been observed to undergo flares (e.g. \citealt{Pacciani2010:ApJ.716L}) and high energy rapid variability. 
In this scenario the emission in a given regime is dominated by synchrotron or inverse Compton (IC) emission from the jet which, because of its angle towards us, is much brighter than the thermal and other emission associated with the AGN. 
Even though the source is close to the Galactic centre, the Galactic column density is relatively low in that direction, allowing the emission to be visible through the Galaxy. The excess column density required by the X-ray spectra suggests excess absorption which may be explained by absorption close to the source, though the apparent excess may also be an effect of spectral curvature. 
The measured IR and X-ray fluxes of this source, approximate radio limits and spectral slope are all broadly consistent with those of blazars in general \citep{Fossati1998:MNRAS.299}. 
In this framework the apparent excess of emission at energies  $>5$\,keV, which we model, phenomenologically, with a second power law is due to the expected inverse Compton emission from blazars (e.g., \citealt{Maraschi1992:ApJ.397}); either IC emission from the interaction of the synchrotron generated photons with the electrons in the jet (synchrotron self Compton; SSC) or from external photons interacting with jet electrons (external-radiation Compton; ERC). However, we cannot constrain this excess component from the \chandra\ spectrum alone and the source is unresolved from the SNR in the higher energy bands (i.e., \swift-BAT, \integral).

Tests of this hypothesis require an optical or IR spectrum with which to confirm the extragalactic nature of the source, though blazars are expected to have no or only very weak lines, so it may be the absence of lines, such as those that would be expected from Galactic sources, that will add weight to the blazar argument. Additionally,  a high level of linear polarisation might  be expected if the source is a blazar, due to the synchrotron emission.  However, the main test is a broader-band SED spanning from radio to optical and into the high energy X-rays,
which should display the double synchrotron and inverse Compton peaks, or at least slopes, and confirm the absence of a thermal component. 
If this source is confirmed as a blazar, at $l,b = 356.81, -1.70$ degrees, while not the first in the Galactic plane \citep{Vandenbroucke2010:ApJ.718}, it will be the first identified so close to the Galactic centre \citep{Massaro2009:A&A.495,Massaro2010:arXiv1006}.


\section{Conclusion}\label{section:conclusion}

On the basis of positional coincidence and common spectral slope, we have identified a new infrared counterpart to the X-ray point source, \cxopt\ in the field of  \igr, visible from 2.2 to 24\,$\mu$m. 
Multi-wavelength spectral modelling shows that the data are consistent with a reddened and absorbed single power law over five orders of magnitude in frequency. This implies non-thermal, possibly synchrotron emission that renders the previous suggestion that this source may be a pulsar, and  its association to the extended SNR emission, unlikely.  We propose that the emission may be due to a blazar viewed through the plane of the Galaxy, and we suggest a number of tests of this hypothesis.


\section*{Acknowledgements}
We thank the referee for their constructive comments. 
This work was supported by the Centre National d'Etudes Spatiales (CNES) and based on observations obtained with MINE: the Multi-wavelength INTEGRAL NEtwork.
JAT acknowledges partial support from Chandra award number GO1-12046X issued by the Chandra X-ray Observatory Center, which is operated by the Smithsonian Astrophysical Observatory for and on behalf of NASA under contract NAS8-03060. 

\label{lastpage}

\end{document}